\documentclass[apj]{emulateapj}

\usepackage{amsmath}
\usepackage{amssymb}
\usepackage{graphicx}
\usepackage{wasysym}
\usepackage{multirow}
\usepackage{mdframed}
\usepackage[breaklinks,colorlinks,citecolor=blue]{hyperref}

\slugcomment{Draft version January 15, 2016, ApJ accepted}

\shorttitle{Helical Magnetic Fields in the NGC1333 IRAS 4A Protostellar Outflows}
\shortauthors{Ching et al.}

\begin{document}


\title{Helical Magnetic Fields in the NGC1333 IRAS 4A Protostellar Outflows}


\author{Tao-Chung Ching\altaffilmark{1,2}, Shih-Ping Lai\altaffilmark{1,3}, Qizhou Zhang\altaffilmark{2}, Louis Yang\altaffilmark{4}, Josep M. Girart\altaffilmark{5,2}, and Ramprasad Rao\altaffilmark{3}}

\email{tching@cfa.harvard.edu}


\altaffiltext{1}{Institute of Astronomy and Department of Physics, National Tsing Hua University, Hsinchu 30013, Taiwan}
\altaffiltext{2}{Harvard-Smithsonian Center for Astrophysics, 60 Garden Street, Cambridge MA 02138, USA}
\altaffiltext{3}{Institute of Astronomy and Astrophysics, Academia Sinica, P.O. Box 23-141, Taipei 10617, Taiwan}
\altaffiltext{4}{Department of Physics and Astronomy, University of California, Los Angeles, California 90095-1547, USA}
\altaffiltext{5}{Institut de Ci\`{e}ncies de l'Espai, (CSIC-IEEC), Campus UAB, Carrer de Can Magrans, S/N, 08193 Cerdanyola del Vall\`es, Catalonia, Spain}
%%\altaffiltext{3}{Institute of Astronomy and Astrophysics, Academia Sinica, 645 N. Aohoku Place, Hilo, HI 96720, USA}


\begin{abstract}
% ===>JMG EDITED 28set15
%
%We present the polarization observations of the CO $J$ = 3--2 line in the 
%NGC1333 IRAS 4A protostellar outflows using the Submillimeter Array (SMA). Our
% CO Stokes $I$ maps at $\sim$ 1$\arcsec$ resolution reveal two independent 
%bipolar outflows from the binary sources of the NGC 1333 IRAS 4A. 
%The kinematic features of the CO  outflows 
We present Submillimeter Array (SMA) polarization observations of the CO $J$ = 3--2 line toward the NGC1333 IRAS 4A. The CO Stokes $I$ maps at an angular resolution of $\sim$1$\arcsec$  reveal two bipolar outflows from the binary sources of the NGC 1333 IRAS 4A. The kinematic features of the CO emission
 % <===JMG EDITED 28set15
can be modeled by wind-driven outflows at $\sim$ 20$\arcdeg$ inclined from the plane of the sky. 
% ===>JMG EDITED 28set15
%The CO $J$ = 3--2 polarization detections at $\sim$ 2$\farcs$3 resolution have polarization directions mostly parallel to the magnetic fields inferred from dust polarizations. 
%The directions of the CO polarization appear to vary smoothly from the hourglass fields of the core to arc-like structures wrapping around the outflow, for the first time revealing
Close to the protostars the CO polarization, at an angular resolution of $\sim$$2\farcs3$, has a position angle approximately parallel to the magnetic field direction inferred from the dust polarizations. The CO polarization direction appears to vary smoothly from an hourglass field around the core to an arc-like morphology wrapping around the outflow, suggesting
 % <===JMG EDITED 28set15
 a helical structure of magnetic fields that inherits the poloidal fields at the launching point and consists of toroidal fields at a farther distance of outflow. The helical magnetic field is consistent with the theoretical expectations for launching and collimating outflows from a magnetized rotating disk. Considering that the CO polarized emission is mainly contributed from the low-velocity and low-resolution data, the helical magnetic field is likely a product of the wind-envelope interaction in the wind-driven outflows. The CO data reveal a PA of $\sim$ 30$\arcdeg$ deflection in the outflows. The variation in the CO polarization angle seems to correlate with the deflections. We speculate that the helical magnetic field contributes to $\sim$ 10$\arcdeg$ deflection of the outflows by means of Lorenz force.
\end{abstract}



\section{Introduction}

% ===>JMG EDITED 28set15
%Magnetic fields are believed to play important roles in the outflow of protostar, 
%as well as star formation itself  \citep{2014prpl.conf..173L, 2014prpl.conf..451F}. 
%The magnetic field in a magnetized accretion disk is expected to be composed 
%of poloidal fields due to the collapse of gas and toroidal fields due to the rotation of disk. 
Magnetic fields are believed to play an important role in the star formation process, and in particular, in the driving of protostellar  outflows  \citep{2014prpl.conf..173L, 2014prpl.conf..451F}. The magnetic field in a magnetized accretion disk is expected to consist of a poloidal component due to the infall motions at the scale of dense cores and a toroidal component due to the rotation at the disk scale. 
% <===JMG EDITED 28set15
Basic MHD theory demonstrates that once the magnetic field reaches a critical angle of $\sim$ 30$\arcdeg$ with respect to the disk plane, the poloidal field may lift the fast-rotating gas off the surface of the disk. The toroidal field may collimate the flow into a bipolar wind. 
% ===>JMG EDITED 28set15
%Composed of poloidal and toroidal fields, a helical structure therefore is 
%expected for the magnetics fields of the winds 
The magnetic field associated with the winds are expected to have a toroidal and poloidal component forming a helical structure \citep{1982MNRAS.199..883B}. Although the exact location where the wind is launched is still under debate (disk wind: \citeauthor{2000prpl.conf..759K} \citeyear{2000prpl.conf..759K}; X-wind: \citeauthor{2000prpl.conf..789S} \citeyear{2000prpl.conf..789S}), there is a general consensus that magnetic fields are essential in launching and collimating the wind.

Magnetic fields and outflows are also important in the formation of protobinary systems. The rotational energy of a protostellar core promotes fragmentation, but the removal of rotational energy by magnetic braking and outflows suppresses fragmentation \citep{2008A&A...477...25H, 2008ApJ...677..327M,  2013ApJ...768..110C}. 
% ===>JMG  28set15: SHOULD BE REPHRASED
Theory predicts that close binaries ($\la$ 10 AU) which fragment in a circumbinary disk may have a common outflow, and wide binaries ($\ga$ 1000 AU) which fragment before the disk formation 
% <===JMG  28set15
are expected to have two independent outflows \citep{2009ApJ...704L..10M, 2009ApJ...706L..46D}. Against this theoretical picture of similar multiplicity of outflows and protostars, it is found that multiple protostellar systems predominantly show single Herbig-Haro objects or protostellar outflows \citep{2000AJ....120.3177R, 2002AJ....124.1045R}. To explain why multiple protostars only have a single jet, \citet{2008A&A...478..453M} modeled the jet interaction from binary protostars and suggested that magnetic field has a significant impact on refocusing jets from two protostars into a single jet.

Polarization observations of thermal dust emission and molecular line emission have been proven to be the most effective method to map the magnetic morphology in molecular clouds \citep{2012ARA&A..50...29C}. Linear polarized dust emission arises from magnetically aligned dust grains, and the direction of polarization is perpendicular to the magnetic field component on the plane of sky \citep{2007JQSRT.106..225L}. Linear polarized spectral line emission arises from the Goldreich-Kylafis effect \citep[][hereafter GK effect]{1981ApJ...243L..75G, 1982ApJ...253..606G, 1983ApJ...267..137K, 1984ApJ...285..126D, 1988ApJ...328..304L}. In the presence of a magnetic field, a molecular rotational level ($J$) split into magnetic sublevels ($M$ = $-J$, $-J$+1, ..., $J$). When the magnetic field is on the plane of the sky, $\Delta$$M$ = $\pm$1 transitions emit linearly polarized radiation with polarization directions perpendicular to the magnetic field, and $\Delta$$M$ = 0 transitions emit linear polarization parallel to the magnetic field. GK effect predicts that due to the different optical depths of the orthogonal polarizations, the net polarization direction is either perpendicular or parallel to the projection of the magnetic field on the plane of the sky. Since the optical depth depends on the anisotropy of the velocity gradient and/or radiation field, the strongest polarization is expected when the anisotropy of velocity gradient and/or radiation field is maximum and when the optical depth is close to unity.  

Compared to the polarization of thermal dust emission, molecular line observations provide kinematic information, and offer possibilities to reveal magnetic field structures in a position-position-velocity space. However, limited by weak polarization signals of the GK effect, molecular line polarization has only been detected in few molecular species and few objects 
(CO: S140, DR21  \citeauthor{1999ApJ...512L.139G} \citeyear{1999ApJ...512L.139G}; NGC 1333 IRAS 4A \citeauthor{1999ApJ...525L.109G} \citeyear{1999ApJ...525L.109G}; NGC 2024 FIR5 \citeauthor{2001ApJ...546L..53G} \citeyear{2001ApJ...546L..53G}; DR21(OH) \citeauthor{2003ApJ...598..392L} \citeyear{2003ApJ...598..392L}, \citeauthor{2005ApJ...628..780C} \citeyear{2005ApJ...628..780C}; G30.79 FIR 10 \citeauthor{2006ApJ...639..965C} \citeyear{2006ApJ...639..965C}; NGC 2071IR \citeauthor{2006ApJ...650..246C} \citeyear{2006ApJ...650..246C}; G34.4+0.23 MM \citeauthor{2008ApJ...676..464C} \citeyear{2008ApJ...676..464C}; IRAS 18089-1732 \citeauthor{2010ApJ...724L.113B} \citeyear{2010ApJ...724L.113B}; M33 \citeauthor{2011Natur.479..499L} \citeyear{2011Natur.479..499L}; IRC +10216 \citeauthor{2012ApJ...751L..20G} \citeyear{2012ApJ...751L..20G}; Orion KL \citeauthor{2013ApJ...764...24H} \citeyear{2013ApJ...764...24H}; SNR IC 443 \citeauthor{2013A&A...558A..45H} \citeyear{2013A&A...558A..45H};
CS: CRL 2688 \citeauthor{1997ApJ...487L..89G} \citeyear{1997ApJ...487L..89G}; IRC +10216 \citeauthor{1997ApJ...487L..89G} \citeyear{1997ApJ...487L..89G}, \citeauthor{2012ApJ...751L..20G} \citeyear{2012ApJ...751L..20G}; DR21(OH) \citeauthor{2008A&A...492..757F} \citeyear{2008A&A...492..757F}; 
HCO$^+$: DR21 \citeauthor{1997ApJ...479..325G} \citeyear{1997ApJ...479..325G};
SiS: IRC +10216 \citeauthor{2012ApJ...751L..20G} \citeyear{2012ApJ...751L..20G};
SiO: HH211 \citeauthor{2014ApJ...797L...9L} \citeyear{2014ApJ...797L...9L}). 
In this work, with the advantage of high angular resolution of SMA, the velocity information of molecular line polarization, and the knowledge of the magnetic fields in the core from dust polarization, we are able to disentangle the magnetic field structure in the outflows of NGC 1333 IRAS 4A.

NGC 1333 IRAS 4A (hereafter IRAS 4A) is a very young low-mass protostellar core with prominent and highly collimated molecular outflows. IRAS 4A contains a protobinary system with the primary source, 4A1, three times more luminous than the secondary, A2 \citep{2000ApJ...529..477L, 2002AJ....124.1045R}. The distance of NGC 1333 cloud is 235 $\pm$ 18 pc from the VLBI parallax measurements of H$_2$O masers \citep{2008PASJ...60...37H}. The magnetic field in the IRAS 4A core mapped with dust polarization exhibits an overall hourglass morphology and a toroidal field in the pseudodisk, consistent with the theoretical expectation of a magnetized rotating core during its collapse phase (\citealt{2006Sci...313..812G, 2008A&A...490L..39G, 2011A&A...535A..44F, 2015ApJ...814L..28C}; Liu et al. submitted; Lai et al. in prep.).  
% ===>JMG EDITED 28set15
IRAS 4A was the first object where the emission of GK effect was mapped. The CO $J$ = 2--1 polarized emission  is detected along the molecular outflow 
% <===JMG EDITED 28set15
\citep{1999ApJ...525L.109G}. While the resolution of the CO $J$ = 2--1 polarization was relatively poor (9\farcs0 $\times$ 6$\farcs$0), in this work we observed the GK effect in CO $J$ = 3--2 to probe the magnetic field structure of IRAS 4A outflows in detail. While the data include the 870 $\mu$m dust polarization, here we focus on the CO $J$ = 3--2 polarization measurements. In Section \ref{sec_obs} we describe our CO $J$ = 3--2 polarization observations. In Section \ref{sec_result} we present the CO $J$ = 3--2 high angular resolution polarization maps. In Section \ref{sec_discussion} we discuss the magnetic field structure in the outflows inferred from our data. Conclusions are presented in Section \ref{sec_summary}.

\section{Observations and Data Reduction}\label{sec_obs}

The polarization observations were carried out with SMA \citep{2004ApJ...616L...1H, 2006PhDT........32M} between 2004 and 2013 in three array configurations in the 345 GHz band, including data in \citet{2006Sci...313..812G}. The CO $J$ = 3--2 line (345.796 GHz) was simultaneously observed with the continuum emission. The phase center was located at the position of NGC1333 IRAS 4A ($\alpha$, $\delta$)$_{\rm J2000}$ = (03$^h$29$^m$10\fs51, +31$\arcdeg$13$\arcmin$31$\farcs$0). Table \ref{obs_table} lists the dates, array configurations, number of antennas, calibrators, and on-source time of the observations. For the observations before 2010, single receivers were used with a total bandwidth of 2 GHz in each sideband. For the observations after 2010, a dual-receiver mode that tuned the 345 and 400 GHz receivers to the same frequency was used. In two tracks of the observations, the C$^{17}$O $J$ = 3--2 line (337.061 GHz) was also covered. The molecular lines were observed in 104~MHz spectral windows with 64, 128, 256, or 512 channels, providing velocity resolutions from 1.4 km s$^{-1}$ to 0.2 km s$^{-1}$. The gain calibrator was 3C84 in all tracks. The bandpass calibrators were also served as the polarization calibrator in each observation. The absolute flux was determined from observations of planets or planetary moons. The typical flux uncertainty in SMA observations was estimated to be $\sim$ 20$\%$.

The basic data reduction of bandpass, time-dependent gain, and flux calibration were carried out using the MIRIAD package for the single receiver data and the IDL MIR package for the dual receiver data. The intrinsic instrumental polarization (i.e., leakage) of the lower side band and the upper side band were calibrated independently with the MIRIAD task GPCAL to an accuracy of 0.1$\%$ \citep{2008SPIE.7020E..2BM} and removed from the data. The SMA polarimetry is optimized at the frequency of CO $J$ = 3--2, and the frequency-dependent leakage in SMA is negligible \citep{2008SPIE.7020E..2BM}. The continuum visibilities were averaged from channels absent of line emission and imaged in the Stokes $I$, $Q$, and $U$ maps. Self-calibration of the Stokes $I$ continuum emission was performed to refine the gain solutions and applied to the Stokes $I$, $Q$, and $U$ visibilities of the continuum, as well as the CO and C$^{17}$O data. Because the CO data were taken with different spectral resolutions, we resampled the channels into an identical resolution of 1.4 km s$^{-1}$ before producing the Stokes $I$, $Q$, and $U$ channel maps. The continuum emission in the molecular line data was removed with the MIRAD task UVLIN. The Stokes $I$, $Q$, and $U$ maps were independently deconvolved using the CLEAN algorithm. Finally, the polarized intensity, position angle, and polarization percentage along with their uncertainties were derived from the Stokes $I$, $Q$, and $U$ maps using the MIRIAD task IMPOL.

The visibility data set spanned baselines from 10 to 210 k$\lambda$, corresponding to scales of 20$\farcs$6 to 1$\farcs$0. To examine sensitivity to spatial structures, different visibility weightings and $uv$-coverages were used during imaging. The continuum polarization map used the entire visibility data to maximize the sensitivity. For the CO data, two different type of maps were made. First, high angular resolution ($\sim 1\arcsec$) maps were obtained by selecting only the visibilities longer than 50 k$\lambda$.  This high resolution map  reveals the complex morphology of the outflow. Second, low angular resolution maps ($\sim 2\arcsec$) were made by selecting visibilities shorter than 70 k$\lambda$. This maximizes the signal-to-noise ratio for the extended polarized emission. Because the C$^{17}$O emission is faint, the C$^{17}$O maps were generated using the subcompact data only. No significant polarization was detected in the C$^{17}$O emission. Table \ref{map_table} lists the mapping parameters of the images presented in this paper: $uv$-coverages, weighting, synthesized beam, and resulting rms noise. The significantly higher rms noise of the continuum Stokes $I$ emission was due to the limited dynamical range of the SMA. With a total on-source time
% ===>JMG EDITED 28set15. it sounds odd to refer to antenna-hours
% of 162.8 antenna-hours, 
 of 24.4 hours, 
 % <===JMG EDITED 28set15
we achieved a sensitivity in Stokes $Q$ and $U$ maps of 80 mJy beam$^{-1}$ for a spectral resolution of 1.4 km s$^{-1}$. In this paper, we set a polarization cutoff level at 3 times the rms noise of the Stokes $Q$ or $U$ maps. The typical measurement uncertainty in the polarization angle for our observation is $\sim$ 6$\arcdeg$. The orientation of position angles follows the International Astronomical Union convention, and is measured from the local direction to the North Pole with positive values increasing towards the east. The definition of inclination angle in this paper is 0$\arcdeg$ on the plane of sky and 90$\arcdeg$ along the line of sight.

\section{Results}\label{sec_result}

\subsection{CO Stokes $I$ Emission}
Figure \ref{fig_chanmap} shows the channel maps of the CO $J$ = 3--2 emission
 % ===>JMG EDITED 28set15
 %of IRAS 4A in low and high angular resolutions. 
 toward IRAS 4A obtained at $1''$ and $2''$ angular resolution. 
  % <===JMG EDITED 28set15
Two pairs of bipolar outflows are seen in the high angular resolution maps. The highest flow velocity channel ($\pm$ 17.6 km s$^{-1}$ with respect to the IRAS 4A systematic velocity) shows an apparent bipolar outflow. The compact blue and red-shifted emission around 4A1 indicates that 
 % ===>JMG EDITED 28set15
%the bipolar outflow are likely launched from 4A1.  The position angle of the outflow is $-$9$\arcdeg$, and
this bipolar outflow is likely launched by this source.  
The initial position angle of the outflow is $-$9$\arcdeg$, but 
% <===JMG EDITED 28set15
the red-shifted gas appears to bend towards the northeast $\sim$ 6$\arcsec$ from 4A1. The channel map of the intermediate velocity ($\pm$ 6.4 km s$^{-1}$) shows two blue-shifted lobes and one red-shifted lobe. The stronger blue lobe along the axis with a postion angle (PA) of $-$9$\arcdeg$ belongs to the 4A1 outflow. The axis of the weaker blue lobe and the red lobe is at a PA of 19$\arcdeg$.  The weaker blue lobe and the red lobe share a common apex at the location of 4A2, suggesting that 4A2 is likely responsible for
 % ===>JMG EDITED 28set15
% the pair of outflows with 19$\arcdeg$ axis. The 19$\arcdeg$ axis of the CO $J$ = 3--2 outflow is consistent with the 18.9$\arcdeg$ axis of the SiO $J$ = 1--0 outflow from 4A2 \citep{2005ApJ...630..976C}. The red-shifted outflow of 4A1 is bended and merged with the red-shifted outflow of 4A2. It is shown in the high velocity channel that the direction of the red-shifted outflow is bended from -9$\arcdeg$ to 19$\arcdeg$ in the northern part,and in the low velocity channel that the flux of the red lobe is close to the sum of the two blue lobes.
 that pair of the lobes.  In addition, this axis is consistent with the  SiO $J$ = 1--0 outflow axis associated with  4A2, which has a PA of 18$\fdg$9 \citep{2005ApJ...630..976C}. The red-shifted outflow of 4A1 appears to be bended and merged with the red-shifted outflow of 4A2. The bending is shown in the high velocity channel, where the direction of the northern, red-shifted lobe changes from a position angle of $-$9$\arcdeg$ to 19$\arcdeg$. The merging is shown in the low velocity channel, where the CO $J$ = 3--2 flux arising from the red lobe is similar to the sum of the flux in the two blue lobes. 
 % <===JMG EDITED 28set15
In the low velocity channel ($\pm$ 6.4 km s$^{-1}$), the red-shifted lobe is tilted by few degrees with respect to the 19$\arcdeg$ axis of the 4A2 outflow. The morphology of the tilt is similar to the wiggle structure of the HCN outflow \citep{2001ApJ...553..219C}, possibly induced by the bending and merging of the 4A1 and 4A2 red-shifted outflows.  
% ===>JMG EDITED 28set15

Another interesting feature is that the
 % ===>JMG EDITED 28set15
 outflows exhibit cone-like structures at low velocities. The sizes of the cones are consistent with the half-maximum contours of the lobes in the low-resolution maps, indicating that the cone-like features arise from the limb-brightened effect of outflow shells. The cones are located further away from the protostars as the flow velocity increases, suggesting that the outflow shells are expanding \citep[e.g. HH212:][]{2015ApJ...805..186L}. Between the 4A1 protostar and the cone's apex, the 4A1 outflow exhibits a  jet-like structure that becomes more prominent in high velocity channels, indicating that the jet-like structure is the high velocity component of the outflow.

Figure \ref{fig_pv} shows the position-velocity (PV) diagrams of the CO emission along the outflow axes of 4A1 and 4A2. The PV diagram of 4A1 shows two types of structures: a slower parabolic structure and a faster linear structure. The parabolic structure is tilted opening up from the star. The edges of parabola are traced by the high-resolution emission, whereas the low angular resolution emission fills out the region enclosed in the parabola. The parabolic structure in the PV diagram is associated with the cone-like structures in the channel maps, since a cut of parabola in the position-velocity space at a fixed velocity plane results in an ellipse in the position-position space. The linear structure has a steeper velocity gradient than the parabolic one. Within 2$\arcsec$ from 4A1, the velocity is rapidly accelerated by 20 km s$^{-1}$. Having very weak emission seen in the low resolution, the linear structure is associated with the fast compact jet-like component of the 4A1 outflow in the channel maps. The PV diagram of 4A2 has a parabolic feature. However, 4A2 does not have a associated fast linear component. 

\subsection{CO and Dust Polarization}\label{sec_3_2}
The CO $J$ = 3--2 polarization is detected both in the 4A1 and 4A2 outflows of the low-resolution maps in Figure \ref{fig_chanmap}. The strongest and most extended polarized emission arises from the region where the red-shifted outflows of 4A1 and 4A2 merge. The polarized intensity is stronger at the edges than the central regions of the outflows, likely because GK effect favors the properties of lower densities and anisotropic optical depth at edges to generate stronger polarization \citep{1988ApJ...328..304L}.  
% ===>JMG EDITED 28set15
%At the same spatial location of the outflows, the polarization detections in 
%different channels are roughly consistent in magnitude and orientation. 
The polarization emission detected in different channels is roughly consistent in magnitude and orientation. 
% <===JMG EDITED 28set15
Because the high-velocity Stokes $I$ emission is weaker, little polarization signals are detected in the high-velocity channels. Figure \ref{fig_spec} shows the Stokes $I$, $Q$, and $U$ spectra at the peaks of the red and blue-shifted polarized emission. The strong Stokes $I$ intensity indicates the CO line is probably optically thick at low velocities. The CO spectra show that the polarized emission 
% ===>JMG EDITED 28set15
arises mainly from the low-velocity channels, 
% <===JMG EDITED 28set15
suggesting that the magnetic fields revealed by CO polarization are associated with the cone-like structures of the outflows.

\begin{figure}
\includegraphics[scale=.65]{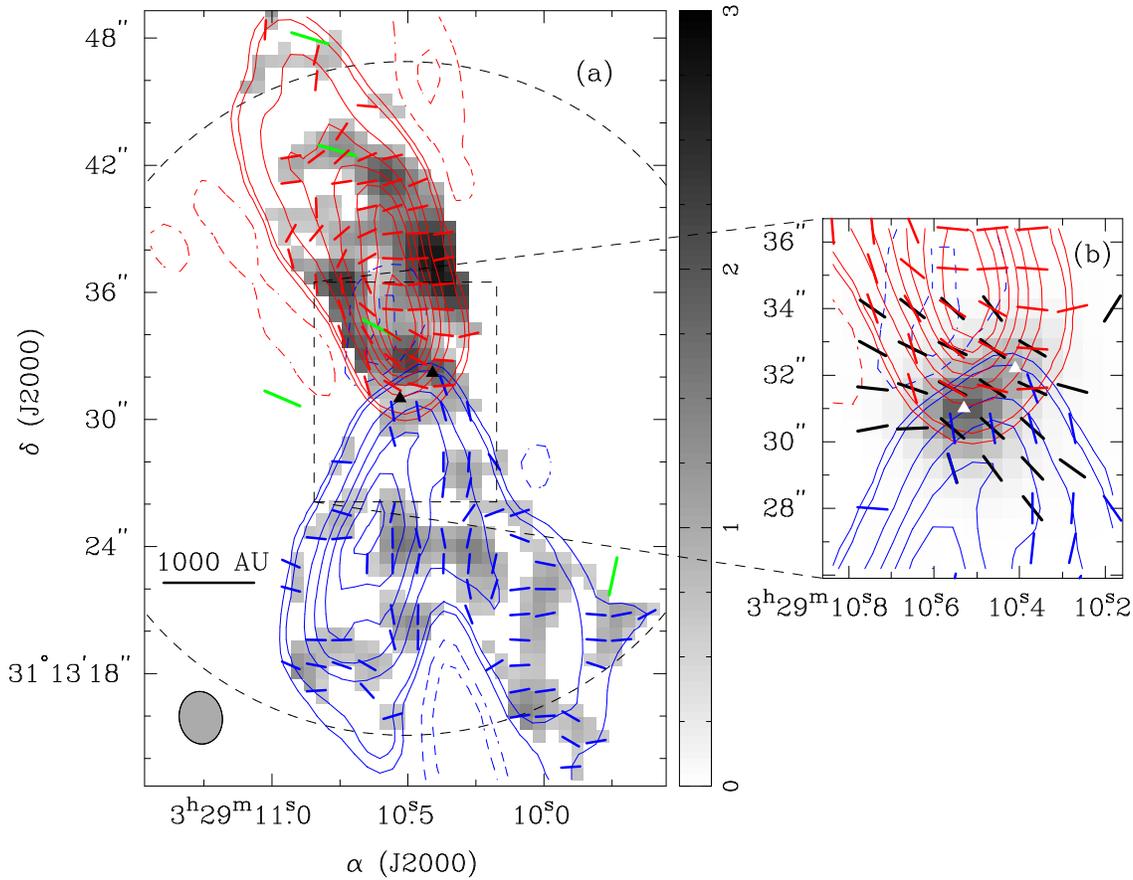}
\caption{A composite map of CO $J$ = 3--2, $J$ = 2--1, and 870 $\mu$m dust polarization maps. The CO $J$ = 3--2 polarization maps integrated over $\Delta$$V$ = $\pm$ 3.5--12.0 km s$^{-1}$ with respect to the systematic velocity of IRAS 4A. The contour levels of CO $J$ = 3--2 emission are $-$10, $-$5, 5, 10, 40, 70, 100, 130, 160 and 190 times $\sigma$ = 0.48 Jy beam$^{-1}$ km s$^{-1}$. The red and blue segments represent the CO $J$ = 3--2 polarizations in the red-shifted and blue-shifted emission. The green segments represent the CO $J$ = 2--1 polarizations \citep{1999ApJ...525L.109G}. The black segments show the magnetic fields inferred from dust polarizations. The length of the segments is unified. (a) CO polarization map of IRAS 4A outflows. The gray scale shows the CO $J$ = 3--2 polarized intensity in units of Jy beam$^{-1}$ km s$^{-1}$. (b) CO polarization map overlapped with the dust polarization map in the central region. The gray scale represents the dust Stokes $I$ emission.}
\label{fig_pol}
\end{figure}

Figure \ref{fig_pol} shows the CO $J$ = 3--2 polarization map integrated over the velocity range of $\Delta$$V$ = $\pm$ 3.5--12.0 km s$^{-1}$, where the polarization emission is the strongest. In the red-shifted lobe, the polarization segments are parallel to the outflow axis at the eastern edge and perpendicular to the outflow axis in the west. The polarization segments in the outflow center show a continuous rotation of the polarization angles connecting the parallel and perpendicular segments. From east to west, the rotation is clockwise in the northern region and counter-clockwise in the southern region. In the western edge of the outflow, the variation of polarization angles from PA of $\sim$ 90$\arcdeg$ in the south to PA $\sim$ 120$\arcdeg$ in the north seems to correlate with the deflection of outflow.

In the blue-shifted lobes, the polarization emission is more scattered. The polarization angles are at PA of $\sim$ 90$\arcdeg$ at the eastern and western edges but change abruptly to PA of $\sim$ 0$\arcdeg$ at the center of the lobes. The 90$\arcdeg$ change in the polarization angle between the center and edges can be caused by the change in magnetic field geometry or by the either-parallel-or-perpendicular orientation of the GK effect. If the change is due to geometrical variations of magnetic fields, extended polarized emission and smooth variation of polarization angles are expected, similar to those in the red-shifted lobe. The existence of gaps between the vertical and horizontal polarizations seems to favor a scenario that the change is caused by the GK effect of CO molecules in different physical conditions.

Figure \ref{fig_pol}b shows the CO polarization segments overlapped with the magnetic field direction inferred from the dust polarization. The uncertainty-weighted mean position angle difference between the CO red-shifted polarizations and the dust polarization is 22$\fdg$9 $\pm$ 1$\fdg$5, indicating that the CO red-shifted polarizations are more likely parallel than perpendicular to the magnetic field in the IRAS 4A core. The CO polarization angles vary smoothly from the IRAS 4A core to the red-sifted lobe at farther distance without any abrupt changes, implying that the CO polarization remains parallel to the magnetic field in the red-shifted lobe. However, the CO polarization segments in the blue-shifted lobes near the protostars show a larger discrepancy with the magnetic fields inferred from dust polarizations. The uncertainty-weighted mean position angle difference between the CO blue-shifted polarizations and the magnetic fields is  33$\fdg$4 $\pm$ 3$\fdg$0, hence it is less clear whether the CO blue-shifted polarizations are parallel or perpendicular to magnetic fields.

\subsection{C$^{17}$O emission}\label{sec_3_3}
\begin{figure}
\includegraphics[scale=.5]{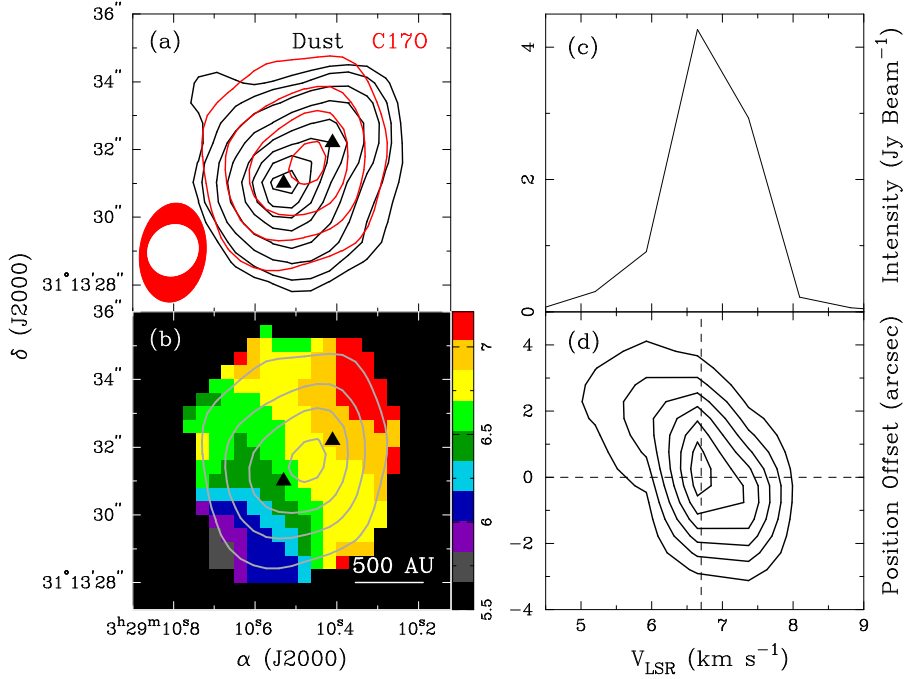}
\caption{Integrated map, intensity-weighted mean $V_{LSR}$ map, spectrum, and PV diagram of C$^{17}$O $J$ = 3--2 emission. Panel (a): Contour maps of the dust (black) and C$^{17}$O (red) integrated emission. The contour levels of dust emission are 3, 5, 10, 20, 40, 60, 80, and 100 times $\sigma$ = 23 mJy beam$^{-1}$. The contour levels of C$^{17}$O emission are 5, 10, 15, and 20 times $\sigma$ = 0.3 Jy beam$^{-1}$ km s$^{-1}$. The synthesized beams of dust and C$^{17}$O data are shown in the left-bottom corner. Panel (b): color image of the C$^{17}$O intensity-weighted mean $V_{LSR}$ (1st moment) overlapped with the contours of the C$^{17}$O integrated emission. The wedge is in units of km s$^{-1}$. Panel (c): C$^{17}$O spectrum at the peak of the integrated emission. Panel (d): C$^{17}$O PV diagram along the position angle from 4A1 to 4A2 ($-$52$\arcdeg$). The contour levels are 5, 10, 15, 20, 25, and 30 times $\sigma$ = 130 mJy beam$^{-1}$. The horizontal and vertical dashed lines label the positions of C$^{17}$O emission peak the and the systematic velocity.}
\label{fig_C17O}
\end{figure}

Figure \ref{fig_C17O}a shows 
% ===>JMG EDITED 28set15
%the integrated emission of the C$^{17}$O J = 3--2 overlapped 
%with the dust emission. The region of the C$^{17}$O emission 
%appears almost the same as the dust emission. 
the C$^{17}$O $J$ = 3--2 integrated emission overlapped with the dust emission.  The  C$^{17}$O emission appears to correlate well with the dust emission. 
Given that the C$^{17}$O emission is optically thin in protostellar cores \citep[it is $\sim$ 1000 times less abundant than the main CO isotopologue,][]{2008MNRAS.383..705C} and it is thermalized \citep[the density of the IRAS 4A core, $\sim$ 10$^7$ cm$^{-3}$, is much higher than the critical density of C$^{17}$O $J$ = 3--2 line, 10$^5$ cm$^{-3}$,][]{2006Sci...313..812G}, the C$^{17}$O $J$ = 3--2 line is a good tracer of the IRAS 4A core. 
% <===JMG EDITED 28set15
The intensity-weighted velocity map (Figure \ref{fig_C17O}b) of the C$^{17}$O line shows a clear velocity gradient along the northwest-southeast direction. The position angle of the maximum velocity gradient is very close to the direction connecting 4A1 and 4A2. The C$^{17}$O spectrum (Figure \ref{fig_C17O}c) shows a single-peak profile, and the PV diagram (Figure \ref{fig_C17O}d) shows a constant rotational velocity of the IRAS 4A core. All these characteristics indicate that the C$^{17}$O $J$ = 3--2 emission mainly reveals a solid-body rotation in the IRAS 4A core with the rotational axis perpendicular to the plane of the 4A1 and 4A2 binary system. The rotation of IRAS 4A core agrees with the velocity pattern of the parent cloud at the few thousands AU scale, as traced by the lower density N$_2$H$^+$ $J$ = 1--0 line \citep{2001ApJ...562..770D}. 

\section{Discussion}\label{sec_discussion}

\subsection{Driving Mechanism of the IRAS 4A CO Outflows}\label{sec_4_1}
The wind-driven model at present is one of the most practical models to study the driving mechanism of outflows \citep[e.g.][]{2000ApJ...542..925L}. In the wind-driven model, molecular outflows are the ambient material swept-up by the wide-angle wind from a young star. The intersection of the wind and the ambient gas determines the shape and velocity of the outflow. For a constant-velocity wind with its mass-loss rate $\propto$ sin$^{-2}\theta$ blowing into an ambient medium with density profile $\propto$ sin$^2\theta$ $r^{-2}$, the outflow has an approximately parabolic shape in which the velocity increases with the distance from the star \citep{1996ApJ...472..211L}. As a result, the shell of the parabolic outflow gives rise to ellipse or parabolic shapes (depending on the inclination angle of outflow) in channel maps and parabolic shapes in PV diagrams.

\begin{figure}
\includegraphics[scale=0.8]{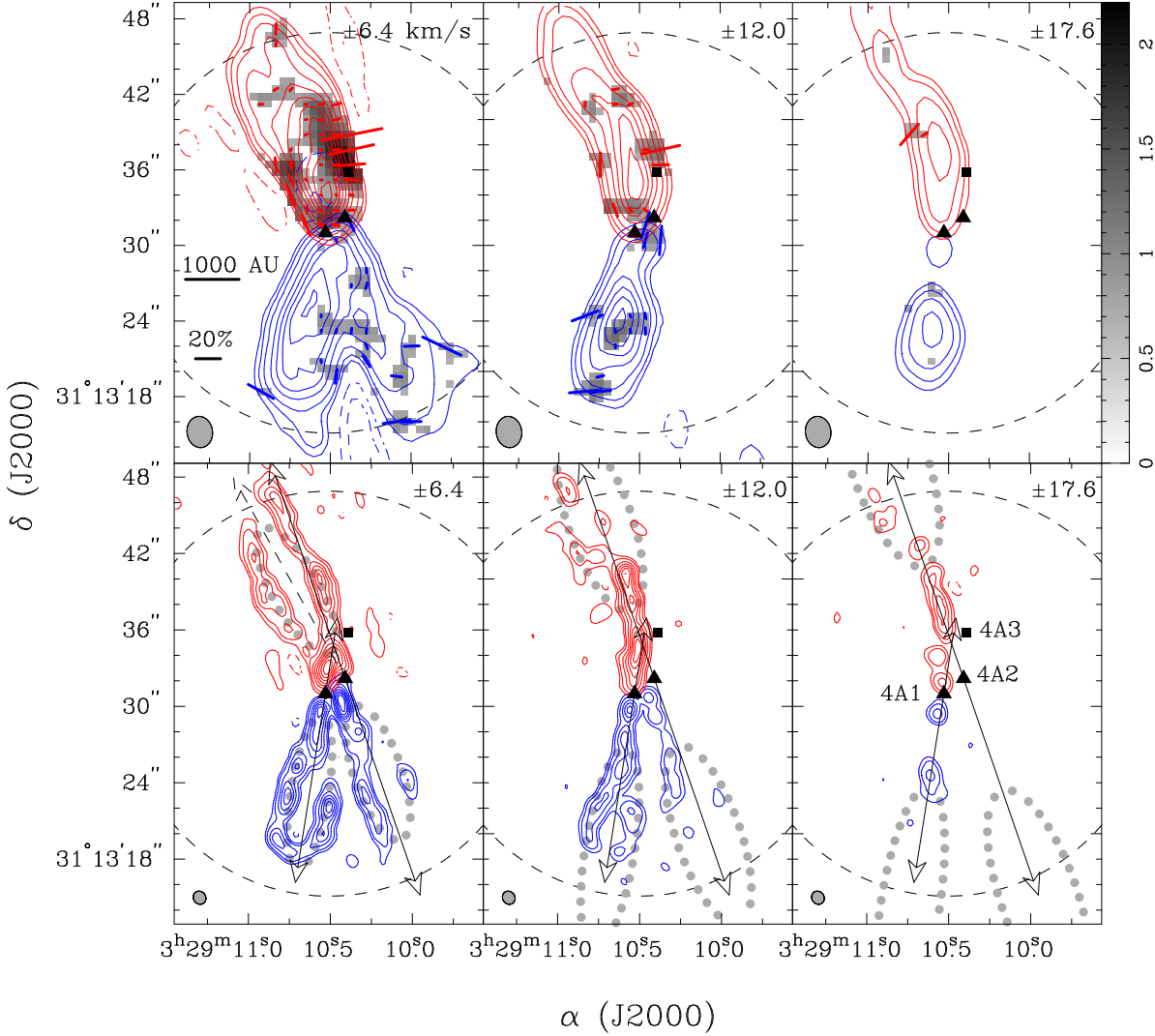}
\caption{ IRAS 4A CO $J$ = 3--2 channel maps integrated over 5.6 km s$^{-1}$ intervals. The top row shows the low-resolution maps, and the bottom row shows the high-resolution maps. The central velocity of each channel with respect to the systematic velocity of $V_{LSR}$ = 6.7 km s$^{-1}$ \citep{2001ApJ...553..219C} are shown in the upper-right corners of the panels. The triangles mark the positions of 4A1 and 4A2. The square marks the 3 mm continuum peak of 4A3 \citep{2015A&A...584A.126S}. The solid-line arrows represent the axis of $-$9$\arcdeg$ in the 4A1 outflows and the axis of 19$\arcdeg$ in 4A2 outflows. The dashed-line arrow in the $\pm$ 6.4 km s$^{-1}$ high-resolution map represents the position angle of 29$\arcdeg$ for the red-shifted wiggle structure. The contour levels of low-resolution and high-resolution maps are $-$10, $-$5, 5, 10, 20, 40, 60, 80, 100, 120, 160, and 180 times $\sigma_{low}$ = 0.45 Jy beam$^{-1}$ km s$^{-1}$ and $-$5, 5, 10, 20, 30, 40, 50, 60, and 70 times $\sigma_{high}$ = 0.15 Jy beam$^{-1}$ km s$^{-1}$, respectively. The red and blue segments represent the polarization detections in the red-shifted and blue-shifted emission. The gray scale shows the CO polarized intensity in units of Jy beam$^{-1}$ km s$^{-1}$. The length of segment is proportional to the polarization percentage. The dotted gray ellipses show the modeled wind-driven outflow shells at the velocities of channels (Section \ref{sec_4_1}). The dashed arcs represent the SMA 31$\farcs$8 primary beam at 345.8 GHz.}
\label{fig_chanmap}
\end{figure}

\begin{figure}
\includegraphics[scale=.6]{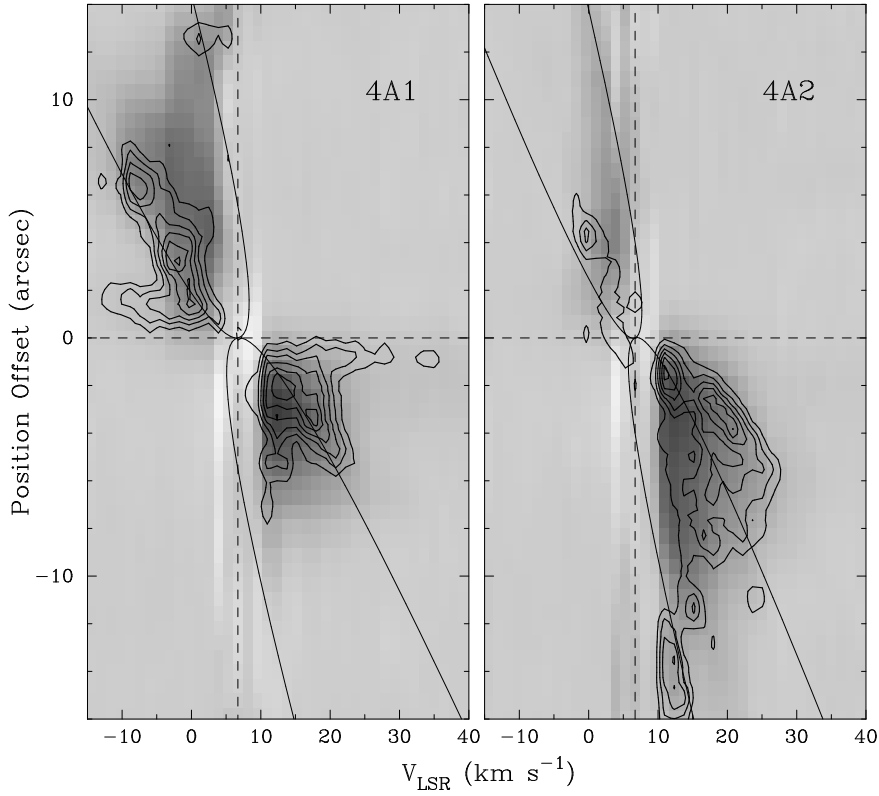}
\caption{PV diagrams of CO $J$ = 3--2 emission along the 4A1 and 4A2 outflow axes (the arrows in Figure \ref{fig_chanmap}). The high-resolution data are represented in contours and the low-resolution data in gray scale. The horizontal and vertical dashed lines label the positions of stars and the systematic velocity. The contour levels are 3, 6, 9, 12, 15, and 18 times $\sigma$ = 0.11 Jy beam$^{-1}$. The curves denote the wind-driven models with $C$ = 3 arcsec$^{-1}$, $v_0$ = 5 km s$^{-1}$ arcsec$^{-1}$, and inclination = 14$\arcdeg$ for the 4A1 outflows and $C$ = 2 arcsec$^{-1}$, $v_0$ = 3 km s$^{-1}$ arcsec$^{-1}$, and inclination = 20$\arcdeg$ for the 4A2 outflows.}
\label{fig_pv}
\end{figure}

Because the observed cone-like and parabolic features in Figures \ref{fig_chanmap} and \ref{fig_pv} are similar to the structures predicted by the wind-driven model, we adopted the simplified analytical model in \citet{2000ApJ...542..925L} to study the driving mechanism of IRAS 4A outflows. In the cylindrical coordinate system, the structure and velocity of the outflow can be written as follows:
\begin{equation}
z = CR^2,\,	v_R = v_0R,\,	v_z = v_0z,
\end{equation}
where $z$ is the distance along the outflow axis; $R$ is the radial size of the outflow perpendicular to $z$; $C$ and $v_0$ are free parameters that describe the spatial and velocity distributions of the outflow shell, respectively. The observed features of 4A1 and 4A2 outflows are reproduced by the models of $C$ = 3 arcsec$^{-1}$, $v_0$ = 5 km s$^{-1}$ arcsec$^{-1}$, and an inclination of 14$\arcdeg$ 
% ===>JMG EDITED 28set15
(with respect to the plane of the sky)
% <===JMG EDITED 28set15
for the 4A1 outflow, and $C$ = 2 arcsec$^{-1}$, $v_0$ = 3 km s$^{-1}$ arcsec$^{-1}$, and an inclination of 20$\arcdeg$ for the 4A2 outflow. The predicted PV structures are shown in Figure \ref{fig_pv}. In the 4A1 and 4A2 PV diagrams, our models show that the blue-shifted parabolic structures are associated with the wind-driven outflows. The displacement between the red-shifted emission and the models is probably caused by the distortion of merging the 4A1 and 4A2 red-shifted outflows. In the 4A1 PV diagram, the fast linear structure cannot be modeled by the wind-driven outflows. The coexistence of a fast linear component and a slow parabolic component in the 4A1 outflows is similar to the HH211 and L1448C outflows \citep{2006ApJ...636L.137P, 2007ApJ...670.1188L, 2010ApJ...717...58H}, in which the fast component is interpreted as a collimated jet, and the slow parabolic component is  modeled as a wind-driven outflow. The coexistence of jet and outflow is consistent with MHD models of protostellar winds \citep[e.g.][]{2006ApJ...649..845S, 2014ApJ...796L..17M}: the highly collimated jet corresponds to the on-axis density enhancement of the wide-opening angle wind, whereas the outflow is mostly consisted of swept-up ambient material. 

The wind-driven model also reproduces the observed cone-like features in the channel maps. Using the same parameters of the PV diagrams, the predicted outflow structures at different velocities are plotted in Figure \ref{fig_chanmap}. In the $\pm$ 6.4 km s$^{-1}$ channel, we adopted an axis of PA=29$\arcdeg$ for the wiggle structure of the red-shifted emission. With the increase of velocity, the morphology of outflow shell changes from a small ellipse close to the protostars to a large ellipse at farther distance. Since the sensitivity falls off toward the edge of the primary beam of the antenna, the ellipse may not be fully recovered and become the observed open structures. Because the red-shifted outflow of 4A1 is bent to the direction of the 4A2 outflow, the predicted red-shifted outflow of 4A1 is not shown in Figure \ref{fig_chanmap}. 

The model provide a rough estimation for 
% ===>JMG EDITED 28set15
%the inclinations of outflows, 
the inclination of the outflow with respect to the plane of the sky, 
% <===JMG EDITED 28set15
since the eccentricity of ellipse in the channel map and the slope of parabola in the PV diagram are determined by inclination. We find that a variation of $\pm$ 5$\arcdeg$ in the inclination angle still results in a reasonable fit to the PV diagrams and the channel maps. We therefore suggest inclinations of 9$\arcdeg$--19$\arcdeg$ for the 4A1 outflow and 15$\arcdeg$--25$\arcdeg$ for the 4A2 outflow. \citet{2011ApJ...728L..34C} suggested an inclination of 10$\fdg$7 by comparing the proper motion from H$_2$ images with the line-of-sight velocity from the SiO spectra. Our finding of 9$\arcdeg$--25$\arcdeg$ is consistent with the picture of the small inclined outflows. The small inclination of the IRAS 4A outflows is also in agreement with the dust polarization pattern, which suggests an inclination of 30$\arcdeg$ for the hourglass-shaped magnetic field in the envelope \citep{2006Sci...313..812G, 2011A&A...535A..44F}. In addition to the wind-driven model, several mechanisms of outflow formation have been used to model the observed position-position-velocity features \citep[e.g.][]{2002ApJ...575..928A}. Although this work does not examine all the models, the fact that the wind-driven model successfully reproduces the observed channel maps and PV diagrams indicates that the 4A1 and 4A2 CO outflows are very likely driven by winds.

\subsection{IRAS 4A Molecular Outflows}
The IRAS 4A outflows have been mapped in several molecular species (e.g. CO: \citeauthor{1995ApJ...441..689B} \citeyear{1995ApJ...441..689B}; \citeauthor{1999ApJ...525L.109G} \citeyear{1999ApJ...525L.109G}; \citeauthor{2012A&A...542A..86Y} \citeyear{2012A&A...542A..86Y}; \citeauthor{2014ApJS..213...13H} \citeyear{2014ApJS..213...13H}; HCN: \citeauthor{2001ApJ...553..219C} \citeyear{2001ApJ...553..219C}; H$_2$CO: \citeauthor{2001ApJ...562..770D} \citeyear{2001ApJ...562..770D} CS: \citeauthor{2001ApJ...562..770D} \citeyear{2001ApJ...562..770D} SiO: \citeauthor{2005ApJ...630..976C} \citeyear{2005ApJ...630..976C}; H$_2$:  \citeauthor{2006ApJ...646.1050C} \citeyear{2006ApJ...646.1050C}). The large-scale maps of CO, SiO, and HCN reveal two 
% ===>JMG EDITED 28set15
%bipolar outflow with few arcmins length in the northeast-southwest (NE-SW) direction and a shorter blue-shifted outflow toward the south. Because the axis of the NE-SW outflow passes right through 4A2, the NE-SW outflows are suggested driven by 4A2, and the shorter blue-shifted southern outflow by 4A1 \citep{2005ApJ...630..976C}.
outflows, one well-collimated bipolar outflow, few arc-minutes in length along the northeast-southwest (NE-SW) direction, and the other, a shorter blue-shifted outflow toward the south. \citet{2005ApJ...630..976C} suggests that the driving source of the NE-SW outflow is 4A2 (the outflow axis  passes right through 4A2) and that 4A1 drives the shorter blue-shifted southern outflow. 
% <===JMG EDITED 28set15
 The BIMA CO maps, however, suggest 4A1 as the driving source since 4A2 is offset from the geometrical center of the CO emission \citep{1999ApJ...525L.109G}. Here our SMA high-resolution maps reveal that both 4A1 and 4A2 launch bipolar outflows: 4A1 launches the CO outflow with a PA of $-$9$\arcdeg$ axis whose blue lobe appears as the shorter southern outflow in the large-scale maps and the red lobe is bent to the direction of the large-scale NE lobe; 4A2 launches the CO outflow with a PA of 19$\arcdeg$ axis as the large-scale NE-SW outflow.

The jet multiplicity of IRAS 4A has also been discovered in the recent work of \citet{2015A&A...584A.126S}. The PdBI SiO, SO, and CO maps at millimeter wavelengths show a C-shaped jet from 4A1 and a mirror-symmetric S-shaped jet from 4A2. Their finding that the jet of 4A1 is faster than the jet of 4A2 is in agreement with our analysis. The S-shaped pattern appears to coincide with the edges of the red and blue-shifted lobes of 4A2 in our maps. The emission of the C-shaped jet is similar to the -9$\arcdeg$ to 19$\arcdeg$ bending of the 4A1 outflow in Figure \ref{fig_chanmap}, although the propagation of the 4A1 jet inferred by \citet{2015A&A...584A.126S} is in the north-south direction, slightly different from our result. Besides 4A1 and 4A2, the 3 mm continuum map shows another peak assigned as 4A3. While the protostellar nature of 4A3 is challenged by the non-detection in other wavelengths, the small displacement between 4A3 and the cross point of 4A1 and 4A2 outflows in Figure \ref{fig_chanmap} seems to suggest a shock-heated origin of 4A3 due to the collision of 4A1 and 4A2 outflows.

In our work, while the designation of the PA=$-$9$\arcdeg$ outflow is clear from the symmetry of the high-velocity gas with respect to 4A1, the association of the PA=19$\arcdeg$ outflow to 4A2 is less straightforward. Is it possible that the 19$\arcdeg$ outflow is driven by the binary system as a whole? In MHD theory, close binaries with a separation of 10 AU may have a common outflow \citep{2009ApJ...704L..10M}, and for binaries with hundreds of AU separation, two independent bipolar outflows are expected \citep{2009ApJ...706L..46D}. Since the separation of 4A1 and 4A2 is $\sim$ 500 AU, it is less likely that the 19$\arcdeg$ outflow is driven by the binary.

\subsection{CO Polarization}
\subsubsection{Linear Polarization}
The CO $J$ = 2--1 polarization detection of the IRAS 4A outflow was presented in \citet{1999ApJ...525L.109G}. Figure \ref{fig_pol}a shows a composite of the CO $J$ = 3--2 and 2--1 polarization maps. The CO $J$ = 2--1 polarizations were detected in a more extended region than the CO $J$ = 3--2. Although the CO $J$ = 2--1 data have on average a relatively large error of $\sim$ 14$\arcdeg$ in polarization angle, the CO $J$ = 2--1 polarization in the center of the red-shifted outflow appears perfectly aligned with the $J$ = 3--2 polarization. At a further distance along the red-shifted outflow, the difference of polarization angles between the two transitions is $\sim$ 70$\arcdeg$. Given the facts that the CO $J$ = 2--1 has lower critical density and more extended emission than those of $J$ = 3--2, the CO $J$ = 2--1 might trace an outer layer of CO outflows and therefore have different polarization angles than the $J$ = 3--2 detections.

\begin{figure}
\includegraphics[scale=.5]{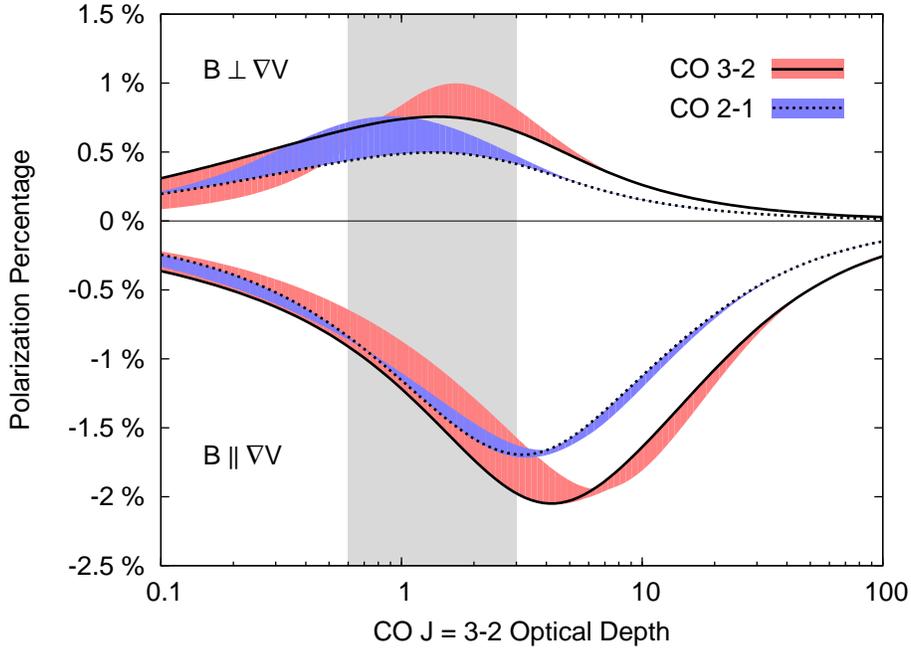}
\caption{Polarization percentage calculated of the CO $J$ = 3--2 (solid lines and red stripes) and $J$ = 2--1 (dashed lines and blue stripes) transitions as a function of the CO $J$ = 3--2 optical depth. The parameters of the solid and dashed lines are $T_{gas}$ = 100 K, $n_{H_2}$ = 10$^4$ cm$^{-3}$, inclination = 20$\arcdeg$, and  $T_{bg}$ = 3 K. Since IRAS 4A has UV radiation of $\sim$ 50 K \citep{2012A&A...542A..86Y}, the red and blue stripes show the variation of the polarization percentage for $T_{bg}$ between 3 to 50 K. Models of magnetic field perpendicular to velocity gradient (B $\bot \, \nabla$V) give the most positive values, and models of magnetic field parallel to velocity gradient (B $\| \, \nabla$V) give the most negative values. Models of magnetic field and velocity gradient oriented between parallel and perpendicular give values between the two extremes. The shadowed region represents the possible range of 0.6--3 optical depth of the IRAS 4A CO $J$ = 3--2 emission.}
\label{fig_gk}
\end{figure}

Following the formulation of \citet{1984ApJ...285..126D} and \citet{2005ApJ...628..780C}, we calculate the CO polarization due to the GK effect. Figure \ref{fig_gk} shows our calculation of the CO $J$ = 3--2 and 2--1 polarization using the properties of gas temperature ($T_{gas}$) of 100 K, number density ($n_{H_2}$) of 10$^{4}$ cm$^{-3}$, background temperature ($T_{bg}$) of 3--50 K \citep{2012A&A...542A..86Y}, and inclination of 20$\arcdeg$ (Section \ref{sec_4_1}) of the IRAS 4A envelope and outflows. The predicted polarization has a range from +1$\%$ to $-$2$\%$ where the positive value means that the polarization is perpendicular to the magnetic field and the negative value means that two are parallel. Because the missing flux in interferometric Stokes $I$ measurement results in higher polarization percentage, the measured CO $J$ = 3--2 polarization reaches $\sim$ 40$\%$ at the edges of the outflows. At the center of the outflows where the missing flux should be minimal, the magnitude of measured polarization is 2.9$\%$, closer to the predicted negative polarization than the positive polarization. The predicated polarization percentage is a function of optical depth and reaches a maximum when optical depth close to unity. The optical depth of CO $J$ = 6--5 emission of IRAS 4A is $\sim$ 3 \citep{2012A&A...542A..86Y}, implying an optical depth of 0.6 for CO $J$ = 3--2 under the LTE condition with $T_{gas}$ = 100 K. Within the optical depth of 0.6--3, the negative polarization is stronger than the positive polarization. Together, our calculation of the GK effect suggests that the orientation of the observed polarization is likely parallel to the magnetic field, consistent with the comparison of the CO and dust polarization (Section \ref{sec_3_2}). 

\begin{figure}
\includegraphics[scale=.65]{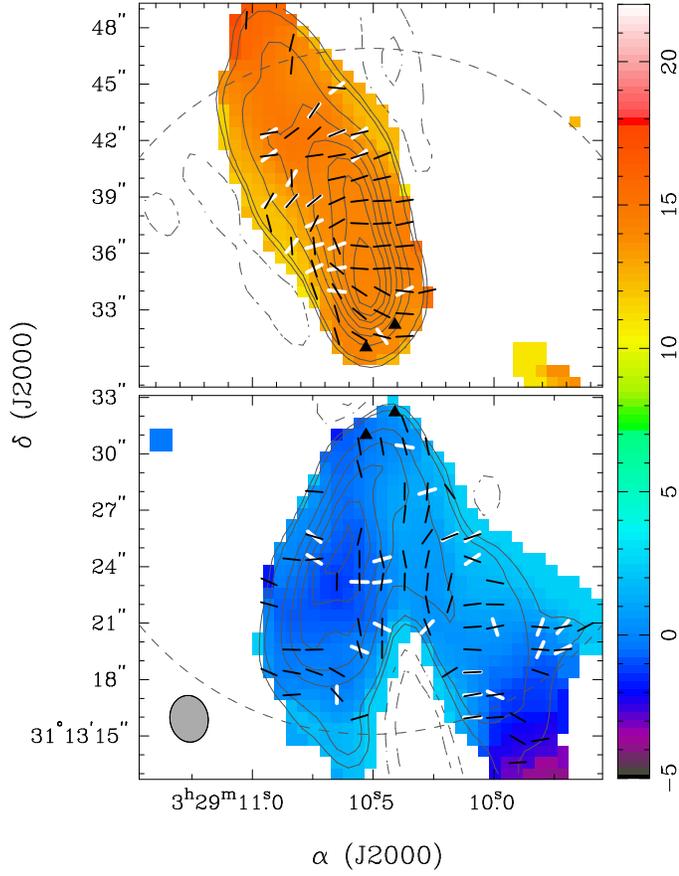}
\caption{Color images of the CO $J$ = 3--2 intensity-weighted mean $V_{LSR}$ (1st moment) overlapped with the direction of velocity gradient (white segments) and CO polarization (black segments) of the red lobe (top panel) and the blue lobe (bottom panel). The mean $V_{LSR}$ and velocity gradient are derived using the data of Figure \ref{fig_pol} ($\Delta$$V$ = $\pm$ 3.5--12.0 km s$^{-1}$ low-resolution data). The white segments show the velocity gradients that are greater than 1.4 km s$^{-1}$ (the velocity resolution of the data) over one synthesized beam. The contours are the same as those in Figure \ref{fig_pol}. The length of the segments is unified and do not reflect the magnitude of polarization nor velocity gradient. The wedge is in units of km s$^{-1}$.}
\label{fig_vg}
\end{figure}

\begin{figure}
\includegraphics[scale=1.5]{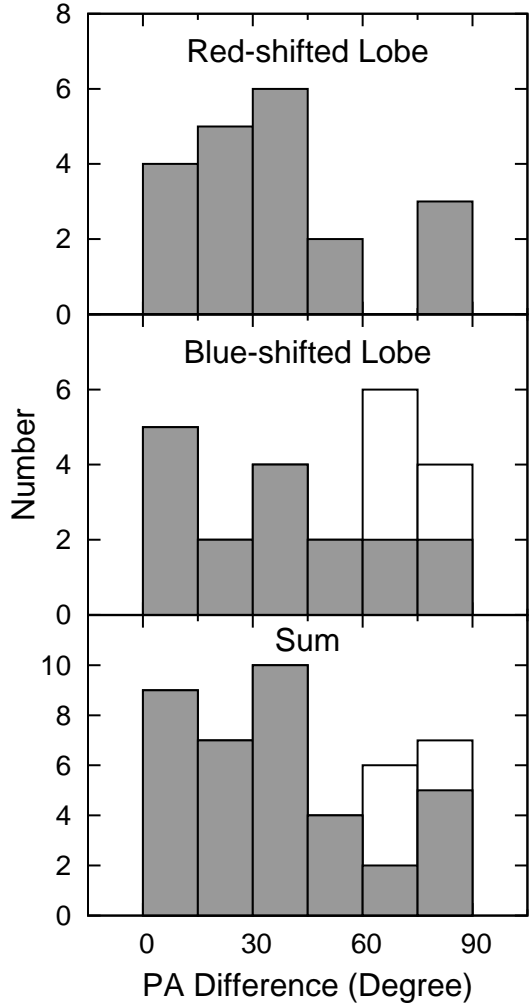}
\caption{Difference between the position angle of the CO $J$ = 3--2 polarization (the black segment in Figure \ref{fig_vg}) and the velocity gradient (the white segment in Figure \ref{fig_vg})  in the red-shifted lobe, the blue-shifted lobe, and the sum of two. The samples from the region where the 4A1 and 4A2 blue-shifted outflows overlap are shown with white boxes.}
\label{fig_hist}
\end{figure}

\citet{2005ApJ...628..780C} pointed out that when radiation field is weak, the anisotropy of the velocity field determines polarization direction of the GK effect. When the velocity gradient is parallel (perpendicular) to the magnetic field, the predicted polarization is parallel (perpendicular) to the magnetic field. Our calculation (Figure \ref{fig_gk}) gives the same results as \citet{2005ApJ...628..780C}. Since the argument implies that the polarization should be always parallel to the velocity gradient, a comparison of the observed CO polarization and velocity gradient provides another examination of GK theory. Figure \ref{fig_vg} shows the velocity field of the CO $J$ = 3--2 emission overlapped with CO polarizations. Figure \ref{fig_hist} shows the histogram of the difference between the polarization angle and the velocity gradient. In the red-shifted lobe, the polarization tends to be parallel to the velocity gradient, in agreement with the theory. In the eastern and western edges of the blue-shifted lobes, the polarization also tends to be parallel to the velocity gradient. But in the central region where the 4A1 and 4A2 blue-shifted outflows overlap, the polarization appears to be misaligned with the velocity gradient. The misalignment might be related to the overlap of the two outflows. Because the observed polarization might be from one outflow and the observed velocity gradient might be from the other, the polarization and velocity gradient are not necessarily aligned when two outflows overlap. 

\subsubsection{Circular Polarization}
Non-Zeeman circular polarization of CO rotational lines has been detected in Orion KL \citep{2013ApJ...764...24H} and SNR IC 443 \citep{2013A&A...558A..45H}. The authors presented a physical model of resonant scattering to explain the observed circular polarization. In the model, the linear polarized emission from a CO population in the background are partially converted to circular polarized emission after scattering off another CO population in the foreground (see also \citeauthor{2014ApJ...795...27H} \citeyear{2014ApJ...795...27H}). As a result of the conversion, the total polarization (Stokes $Q^2$ + $U^2$ + $V^2$) is conserved, but the PA ($\onehalf\tan^{-1}\frac{U}{Q}$) of linear polarization can be changed depending on how much the Stokes $Q$ and $U$ converts to $V$. Therefore, a verification of Stokes $V$ is recommended in the measurement of GK effect in order to give correct interpretation of the PA of linear polarization.

\begin{figure}
\includegraphics[scale=1.5]{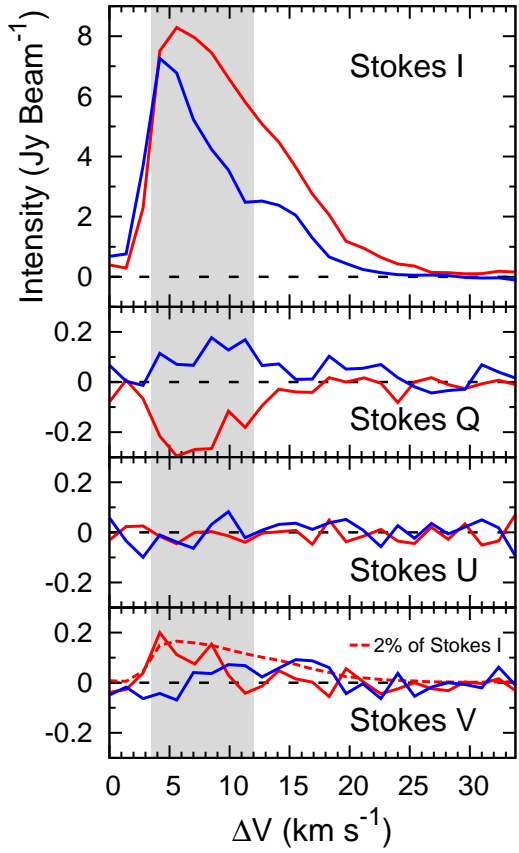}
\caption{CO $J$ = 3--2 Stokes $I$, $Q$, $U$, and $V$ spectra of low-resolution data at the red-shifted (red profiles) and blue-shifted (blue profiles) polarization peaks. The $x$-axis is the velocity with respect to the systematic velocity. The shadowed region represents the $\Delta$$V$ = $\pm$ 3.5--12.0 km s$^{-1}$ integration range of Figure \ref{fig_pol}.}
\label{fig_spec}
\end{figure}

Figure \ref{fig_spec} shows the Stokes $V$ spectrum of the CO $J$ = 3--2 emission in the IRAS 4A.  The CO Stokes $V$ is about 2\% of the Stokes $I$ emission, which is similar to the value measured in the Stokes $V$ dust continuum of IRAS 4A. The dust continuum emission should not show any significant circular polarization. Therefore, the detected circular polarization in both the CO and dust emission is likely dominated by calibration errors. For the SMA polarimeter, because the Stokes $V$ is measured via $V$ = \onehalf ($g_{RR} RR - g_{LL} LL$) where $RR$ and $LL$ are the correlations of right ($R$) and left ($L$) hand circular-polarized feeds and $g_{RR}$ and $g_{LL}$ are the gain solutions, the sensitivity in Stokes $V$ is mainly determined by the uncertainty in the gain solution and the percentage of circular polarization of the gain calibrator. In our gain observations, the noise level of 3C84 was 1--2\% of the flux and we assumed that 3C84 was circularly unpolarized. The noise level of Stokes $V$ in our data therefore is few percent of Stokes $I$. The Stokes $V$ in Figure \ref{fig_spec} is at the same level as the noise, indicating that the percentage of true circular polarization in CO $J$ = 3--2 emission has a conservative upper limit of 2\%.

Granted that the Stokes $V$ spectrum in Figure \ref{fig_spec} is true detection, the property of polarization conservation of resonant scattering allows us to convert the circular polarization back to linear polarization and discuss the initial PA of the GK effect. In Figure \ref{fig_spec}, the integrated Stokes $Q$, $U$, $V$ in the redshifted wing (within the velocity interval showed by the grey bar) are ($Q$, $U$, $V$) = (-1.88, -0.16, 0.74) Jy Beam$^{-1}$ km s$^{-1}$. The PA derived from these $Q$ and $U$ values is -88$\arcdeg$.  Assuming that Stokes $V$ comes all from Stokes $Q$, then the initial linear polarization would be ($Q$, $U$) = (-2.02, -0.16). Reversely,  assuming that all the Stokes $V$ comes from Stokes $U$, then the initial linear polarization would be ($Q$, $U$) = (-1.88, -0.76). These two extreme cases would yield an initial PA of -88$\arcdeg$ and -79$\arcdeg$, respectively.  Therefore, by not including the possible effect of the conversion to circular polarization, we expect a deviation of no more than of 9$\arcdeg$, which is close to the typical uncertainty in the polarization angle and is not significant compared to the 90$\arcdeg$ rotation of the CO red-shifted polarizations from the eastern side to the western side. Therefore, the upper limit of 2\% Stokes $V$ would not change our finding of helical magnetic fields in the outflows.

\subsection{Magnetic Fields in IRAS 4A Outflows}\label{sec_4_4}
Is the magnetic field revealed by the CO polarization the ambient field of dusty envelope, the field of winds, or the field of wind-envelope interaction? We argue that the magnetic field probed by the CO polarization of IRAS 4A is associated with the wind-envelope interaction for three reasons. First, the orientation of the CO polarization is significantly different from the magnetic field in the core. The CSO 350 $\mu$m dust polarization in 20$\arcsec$ resolution reveals that the mean field direction in the IRAS 4A core is 45$\fdg$9 $\pm$ 13$\fdg$6 \citep{2009ApJ...702.1584A}. At the scales of 1$\arcsec$, the SMA 870 $\mu$m dust polarization reveals an hourglass structure of magnetic fields \citep{2006Sci...313..812G}. Since the CO polarization in the outflow is neither 45$\fdg$9 oriented nor hourglass-like, the CO polarization should not reveal the ambient field of the envelope. Second, the polarized CO emission is mainly contributed from the low-velocity and low-resolution data. In MHD theory, winds are expected to be narrow and highly collimated with velocities of order 100--1000 km s$^{-1}$. The fact that the CO outflows in the polarization map have a velocity of $\sim$ 10 km s$^{-1}$, a width of $\sim$ 700 AU (3$\arcsec$ in diameter), and a bent morphology indicates that the CO polarization should not reveal the magnetic field of the wind. Third, the orientation of CO polarizations correlates with the morphology of the outflow. The IRAS 4A CO outflows likely trace shells of ambient material swept-up by a wide-angle wind (Section \ref{sec_4_1}). Given that the variation of polarization angles in the western edge of the red-shifted outflow correlates with the deflection of outflow, the magnetic field revealed by CO polarization is likely associated with the outflow itself, and therefore the field of wind-envelope interaction.

From the base to the northern part of the IRAS 4A red-shifted outflow, the difference between the position angles of the CO polarization and the magnetic field inferred from dust polarization increases from 4$\fdg$3 to 54$\fdg$5 (Figure \ref{fig_pol}b), indicating that the magnetic field of outflow is similar to the parental fields in the core at the launching point and deviate at a farther distance of outflow. In the red-shifted outflow, the CO polarization segments appear in arcs wrapping around the outflow. The morphology of the arc appears to be consistent with the projection of a toroidal field in an inclined outflow. Overall, the CO red-shifted polarizations reveal a helical structure of magnetic fields in outflow that inherits the poloidal field of a hourglass structure at the launching point and becomes toroidal at a farther distance of the outflow. This is in agreement with the expectation of MHD theory.  Helical magnetic fields in jets have been inferred from the linear polarization in the synchrotron emission in HH 80--81 \citep{2010Sci...330.1209C} and SiO polarization in HH211 \citep{2014ApJ...797L...9L}. 
% ===>JMG EDITED 28set15: IT SHOULD BE REWRITTEN, 
%        I DONT FULLY UNDERSTAND THE FIRST SENTENCE
Compared to their detections of few hundred AU away from the driving sources, our CO polarization emission is extended from the core to the outflow, directly tracing the helical structure of magnetic fields from the driving sources to outflows. 

\subsection{Bend of the Red-shifted Outflow}\label{sec_4_5}
The bend of the 4A1 red-shifted outflow from a PA of $-$9$\arcdeg$ to 19$\arcdeg$ is significant. 
Note that this bend in the CO emission at a distance of 6$\arcsec$ from IRAS 4A is different from the PA of 34$\arcdeg$ bend in the SiO red-shifted outflow at a distance of 23$\arcsec$ \citep{2005ApJ...630..976C}. The bend in our map also exists in the CO $J$ = 2--1 outflow \citep{1999ApJ...525L.109G, 2014ApJS..213...13H}. Following \citet{1999A&A...343..558H} theoretical work, \citet{1999ApJ...525L.109G} proposed a scenario that magnetic fields are responsible for the bend. Here we study the bending mechanism in detail with our updated results.

The bending of outflows could be due to the orbital motion of the launching sources in a binary system, the dynamical pressure from external medium, or the Lorenz forces between the current carrying outflows and an external magnetic field \citep[e.g.][]{1998A&A...334..750F}. Since the orbital motion of a binary system gives rise to symmetric bending in both sides of bipolar outflow, it cannot explain the asymmetric bending of the 4A1 red-shifted outflow. The dynamical pressure mechanism requires an external source to collide and deflect outflows. The deflection of the 4A1 red-shifted outflow from the axis of 4A1 to the axis of 4A2 suggests that the collision of the 4A1 and 4A2 red-shifted outflows could be the source of dynamical pressure. Replacing $n_{ism}/n_{jet}$ and $v_{\star}/v_{jet}$ in Equation 5 of \citet{1998A&A...334..750F} by $n_{4A2}/n_{4A1}$ and $v_{4A1-4A2}/v_{4A1}$, the equation can be written as
%\begin{equation}
\begin{multline}
\tan \alpha = 10^{-3} \left(\frac{L_{jet}/R_{jet}}{20}\right) \left(\frac{n_{4A2}/n_{4A1}}{1}\right) \\ \left(\frac{v_{4A1-4A2}/v_{4A1}}{0.01}\right)^2 ,
\end{multline}
%\end{equation}
where $\alpha$ is the deflection angle, $L_{jet}$ is the length scale of the curved trajectory, $R_{jet}$ is the jet radius, $n_{4A2}$ and $n_{4A1}$ are the volume densities of 4A1 and 4A2 outflows, $v_{4A1-4A2}$ is the respective velocity between 4A1 and 4A2 outflows, and $v_{4A1}$ is the velocity of the 4A1 outflow. Modified with the $\theta \sim 20\arcdeg$ inclination of the IRAS 4A outflows, the $L_{jet}$ of 10$\farcs$6/$\cos \theta$, $R_{jet}$ of 1$\farcs$6, $n_{4A2}/n_{4A1} \simeq I_{4A2}/I_{4A1}$ (the intensity ratio of 4A1 and 4A2 outflows) = 0.25, $v_{4A1-4A2}$ of 10 km s$^{-1}$/$\cos \theta$, and $v_{4A1}$ of 10 km s$^{-1}$/$\sin \theta$ give a deflection of $\sim$ 7$\arcdeg$ by the collision of 4A1 and 4A2 outflows. 

Lorentz force might arise if the magnetic field of one outflow acts on the current of the other outflow. The Equation 11 of \citet{1998A&A...334..750F} gives the bending due to Lorentz force
%\begin{equation}
\begin{multline}
\tan \alpha = 0.018 \left(\frac{I_{jet}}{10^{11}\rm\,A}\right) \left(\frac{B_{ext}}{10\,\mu\rm G}\right) \left(\frac{R_{jet}}{10^{15}\rm\,cm}\right)^{-2} \\ \left(\frac{n_{jet}}{100\rm\,cm^{-3}}\right)^{-1} \left(\frac{v_{jet}}{300\rm\,km\,s^{-1}}\right)^{-1},
%\end{equation}
\end{multline}
where $I_{jet}$ is the current carried by a jet, $B_{ext}$ is the strength of external magnetic field, $n_{jet}$ and $v_{jet}$ are the volume density and the velocity of the jet. 
Scaling the typical $I_{jet}$ of 10$^{11}$ A  for $R_{jet} = 10^{15} \, \rm cm, n_{jet} = 100 \, \rm cm^{-3}, \rm and\, v_{jet} = 300 \, \rm km\,s^{-1}$ jets \citep{1995A&A...300..791F}, the $I_{jet}$ of IRAS 4A outflows is $\sim$ 10$^{15}$ A.
Given that $n_{jet} = N_{jet} \cos \theta/2R_{jet}$ ($N_{jet}$ is the column density of outflow $\sim 10^{22}$ cm$^{-2}$; \citeauthor{2012A&A...542A..86Y} \citeyear{2012A&A...542A..86Y}) and the $v_{jet} \sim$ 10 km s$^{-1}/ \sin \theta$, a deflection of 10$\arcdeg$ due to Lorentz force requires magnetic fields of order 10--100 $\mu$G. To date, the only one direct measurement of magnetic field strength in jets of protostars gives a value of 200 $\mu$G \citep{2010Sci...330.1209C}. Considering the magnetic field strength is 1--5 mG in the IRAS 4A core \citep{1999ApJ...525L.109G, 2009ApJ...702.1584A}, 10--100 $\mu$G magnetic fields in the outflow with wind-envelope interaction is reasonable. Therefore, the bend in the IRAS 4A red-shifted outflow could be a combination of collision of the 4A1 and 4A2 outflows and deflection by magnetic fields.

\subsection{IRAS 4A Binary System and Two Bipolar Outflows}
\begin{figure}
\includegraphics[scale=.7]{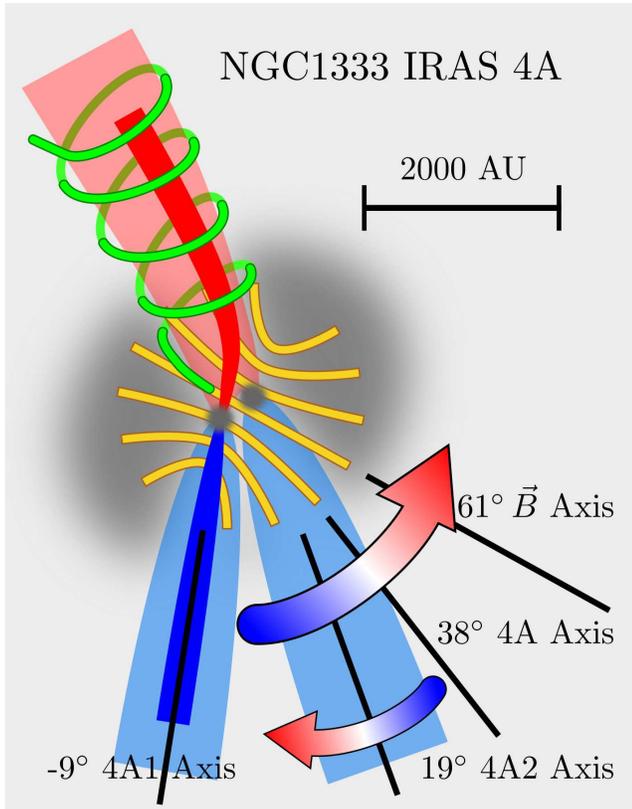}
\caption{Schematic view of the magnetic fields and kinematics in IRAS 4A. The binary, 4A1 and 4A2, of the IRAS 4A (dark grey ellipse) has an axis of 38$\arcdeg$, rotating from east to west (Section \ref{sec_3_3}) and threaded by a 61$\arcdeg$ tilted hourglass magnetic field \citep[yellow lines;][]{2006Sci...313..812G}. The binary launches two pairs of the CO $J$ = 3--2 cone-like outflows (light red and light blue lobes). The rotation of 4A2 is west-to-east with an axis of 19$\arcdeg$ \citep{2011ApJ...728L..34C}. The axis of the jet-like outflows (red and blue stripes) of 4A1 is $-$9$\arcdeg$ and is bended to the axis of 4A2 in the red-shifted side. The outflows are wrapping by a helical structure of magnetic fields (green line). Please note that the green line illustrates the wrapping only and neither represents the hoops of nor the direction of the wrapping.}
\label{fig_cartoon}
\end{figure}

We summarize the structure of the magnetic field, the outflow, and the kinematics of the IRAS 4A binary system in Figure \ref{fig_cartoon}. The observed structures are similar to those predicted in star formation theory. The magnetic field of the core is dragged by gravitational collapse into an hourglass morphology \citep{1987ARA&A..25...23S, 1991ApJ...373..169M}, and the magnetic field of the outflow follows the poloidal component of the hourglass structure at the launching point and become toroidal at large distance \citep{1982MNRAS.199..883B}. The helical magnetic fields of the outflows can help the merging of the two outflows into one \citep{2008A&A...478..453M}. While the mass ratio of the binary tends to be unity because the accretion of rotating gas favors the minor object with faster rotation \citep{2002MNRAS.336..705B}, unequal-mass binary is suggested when the angular momentum of accreting material is moderately removed by magnetic fields \citep{2009ApJ...706L..46D, 2013ApJ...763....7Z}. The misalignment of the magnetic field and the rotation is also a well known fact to reduce the efficiency of magnetic braking \citep{2009A&A...506L..29H, 2012A&A...543A.128J, 2013ApJ...774...82L}. To summarize, our observations are consistent with the theory of protobinary formation that weakly magnetized core fragments due to rotation, and evolves into a binary system separated by hundreds of AU with two independent bipolar outflows \citep{2008A&A...477...25H, 2008ApJ...677..327M, 2009ApJ...706L..46D}. 

Interestingly, the west-to-east rotation of the 4A core and its parent cloud (Section \ref{sec_3_3}) is opposite to the east-to-west rotation of 4A2 \citep{2011ApJ...728L..34C}. This opposite rotation has not been predicted in the theory of protobinary formation from magnetized core. Because magnetic braking cannot turn the rotation around, the opposite rotation of the star to the core seems to be incompatible if the magnetic field is the only source regulating angular momentum. Turbulent fragmentation \citep{2010ApJ...725.1485O} can regulate angular momentum and may be a solution of opposite rotation, and truly, the misalignment of the $-$9$\arcdeg$ 4A1 axis and the 19$\arcdeg$ 4A2 axis is expected in turbulent fragmentation. 

\section{Conclusions}\label{sec_summary}
We present SMA polarization observations of the NGC 1333 IRAS 4A CO $J$ = 3--2 protostellar outflows. With the highest spectral polarization sensitivity to date, we are able to reveal the helical structure of magnetic fields from the protostellar core to the outflow. Our main conclusions are the following:
\begin{enumerate}

\item Two pairs of bipolar outflows are launched independently from the binary sources of IRAS 4A. 4A1 launches the CO outflow with an axis of $\simeq -9\arcdeg$. In large-scale maps, the blue lobe of 4A1 outflow appears as the southern outflow and the red lobe is bended to the NE direction. 4A2 launches the CO outflow with an axis of $\simeq 20\arcdeg$, which agrees with the large-scale NE-SW outflow. 

\item Both the 4A1 and 4A2 outflows exhibit cone-like features in the channel maps and parabolic features in the PV diagrams, which can be modeled with the wind-driven outflow. The 4A1 outflow has an additional jet-like feature in the channel maps and a fast linear feature in the PV diagrams, which can be interpreted as a collimated jet. The coexistence of jet and outflow is consistent with protostellar wind MHD models. The wind-driven model provides a rough estimation of $\sim$ 20$\arcdeg$ inclination for the IRAS 4A outflows, in agreement with the picture of small inclined outflows from the previous works.

\item The CO $J$ = 3--2 linearly polarized emission is strong and is spatially extended in the red-shifted outflow. The comparison between the CO and dust polarizations suggests that the CO polarization is parallel to the magnetic field in the red-shifted lobe. We calculate the predicted CO polarization based on the properties of the IRAS 4A outflows, and the magnitude as well as orientation of the CO red-shifted polarizations are in agreement with the prediction of the GK effect. The CO $J$ = 3--2 blue-shifted polarized emission is weaker than the red-shifted emission, indicating that the blue-shifted polarizations might arise from more complex physical conditions.

\item The magnetic field inferred from the CO $J$ = 3--2 polarizations appear parallel to the hourglass fields of the core and become arc-like wrapping around the outflow, suggesting a helical structure of magnetic fields that inherits the poloidal field at the launching point and consists of a toroidal field at a farther distance of outflow. Belonging to the cone-like components of the outflows, the helical magnetic field is likely a product of the wind-envelope interaction in the wind-driven outflows. The helical magnetic field is in agreement with MHD theory for launching and collimating outflows from a magnetized rotating disk.

\item We examined the circular polarization in our data. The Stokes $V$ in the CO and dust emission is both about 2\% of the Stokes $I$ emission. The 2\% Stokes $V$ is at the same level as the gain calibration errors of the SMA. For the CO polarization, although the upper limit of 2\% Stokes $V$ would induce a deviation of 9$\arcdeg$ in the PA of linear polarization, it is not significant compared to the 90$\arcdeg$ rotation of the CO polarizations from the east to the west of the red-shifted outflow. Our finding of helical magnetic fields in the outflows hence would not be affected by the Stokes $V$ emission.

\item 
The bend of the 4A1 red-shifted outflow from $-$9$\arcdeg$ to 19$\arcdeg$ is significant. While the collision of 4A1 and 4A2 outflows and the Lorentz force between the outflows can contribute to a deflection of $\sim$ 10$\arcdeg$, the bend in the IRAS 4A red-shifted outflow might result from a combination of both bending mechanisms. Our findings yield the first observational evidence that magnetic fields bend and merge outflows.
\end{enumerate}

\acknowledgments
T. C. C. acknowledges the support of the Smithsonian Predoctoral Fellowship and  ALMA-Taiwan Graduate Fellowship. 
T. C. C. and S. P. L. thank the support of the Ministry of Science and Technology (MoST) of Taiwan with Grants NSC 98-2112-M-007-007-MY3, NSC 101-2119-M-007-004 and MoST 102-2119-M-007-004-MY3.
Q. Z. acknowledges the support of the SI CGPS award Magnetic Fields and Massive Star Formation. 
J. M. G. acknowledges the support from MICINN (Spain) AYA2014-57369-C3 grant and the MECD (Spain) PRX15/00435 travel grant.
The Submillimeter Array is a joint project between the Smithsonian Astrophysical Observatory and the Academia Sinica Institute of Astronomy and Astrophysics and is funded by the Smithsonian Institution and the Academia Sinica.

\clearpage
	
%\bibliography{COpaper_150907_revise}
%\bibliographystyle{apj}

%%\nocite{*}

\clearpage

%\begin{landscape}
\begin{deluxetable}{cccccccccc}
\tabletypesize{\scriptsize}
\tablecaption{Observational Parameters\label{obs_table}}
\tablewidth{0pt}
\tablehead{
\multirow{2}{*}{Date} & \multirow{2}{*}{Configuration} & Number of & \multicolumn{2}{c}{Velocity Resolution} & On-source & Polarimeter & \multicolumn{3}{c}{Calibrators} \\\cline{4-5}\cline{8-10} & & Antennas & CO & C$^{17}$O & Observing Time & Mode & Gain & Flux & Bandpass }
\startdata
2004 Dec 05 & Compact & 6 & 0.35 km s$^{-1}$ & -- & 3.5 hr & Single & 3C84 & Ganymede & 3C279 \\
2004 Dec 06 & Compact & 6 & 0.35 km s$^{-1}$ & -- & 6.9 hr & Single & 3C84 & Ganymede & 3C279 \\
2009 Jan 27 & Subcompact & 7 & 0.70 km s$^{-1}$ & 0.72 km s$^{-1}$ & 3.7 hr & Single & 3C84 & Saturn & 3C273 \\
2009 Feb 24 & Extended & 8 & 0.70 km s$^{-1}$ & 0.72 km s$^{-1}$ & 3.6 hr & Single & 3C84 & Ceres & 3C273 \\
2011 Oct 20 & Compact & 7 & 1.41 km s$^{-1}$ & -- & 5.5 hr & Dual & 3C84 & Callisto & 3C454.3 \\
2013 Sep 06 & Subcompact & 6 & 0.18 km s$^{-1}$ & -- & 1.2 hr & Dual & 3C84 & Callisto & 3C111 \\
\enddata
\end{deluxetable}
%\end{landscape}
%\clearpage

\begin{deluxetable}{ccccccc}
\tabletypesize{\scriptsize}
\tablecaption{Mapping Parameters\label{map_table}}
\tablewidth{0pt}
\tablehead{
\multirow{2}{*}{Map} & \multicolumn{2}{c}{$u$, $v$} & \multicolumn{2}{c}{Synthesized Beam} & \multicolumn{2}{c}{rms Noise (mJy beam$^{-1}$)} \\
\cline{2-3} \cline{4-5} \cline{6-7} & Range (k$\lambda$)& Weighting & HPBW(\arcsec) & P.A. (\arcdeg) & Stokes $I$ & Pol}
		\startdata
		Continuum & 10--210 & Uniform & 1.66 $\times$ 1.50 & $-$56.9 & 23 & 2.4 \\
		CO High-resolution & 50--210 & Robust = 1 & 1.07 $\times$ 0.96 & 33.9 & 110 & -- \\
		CO Polarization & 10--70 & Robust = 1 & 2.49 $\times$ 2.03 & 8.2 & 400 & 80 \\
		C$^{17}$O & 10--80 & Natural & 3.00 $\times$ 2.12 & $-$8.33 & 130 & -- \\
		\enddata
\end{deluxetable}

\clearpage

\clearpage

\clearpage

\clearpage

\clearpage

\clearpage

\clearpage

\clearpage

\clearpage

\clearpage

\end{document}